\documentclass{ws-mpla}
\usepackage[super]{cite}
\usepackage{graphicx}
\usepackage{amsmath,amssymb}
\usepackage{color}
\begin{document}

\markboth{Y. Yang et al.}
{Aharonov-Bohm effect on the generalized DKP oscillator in the Som-Raychaudhuri space-time}

\catchline{}{}{}{}{}

\title{Aharonov-Bohm effect on the generalized Duffin-Kemmer-Petiau oscillator in the Som-Raychaudhuri space-time
}

\author{Yi Yang\footnote{yangyigz@yeah.net}, Zheng-Wen Long\footnote{zwlong@gzu.edu.cn (Corresponding author)}, Hao Chen, Zi-Long Zhao and Chao-Yun Long}
\address{College of Physics, Guizhou University, Guiyang, 550025, China}


\maketitle


\begin{abstract}
The generalized Duffin-Kemmer-Petiau (DKP) oscillator with electromagnetic interactions in the curved space-times are investigated. We introduce firstly the generalized DKP oscillator in Som-Raychaudhuri space-time with Cornell potential. Then, we consider the electromagnetic interactions into the generalized DKP oscillator. The energy eigenvalues and eigenfunction of our problem are obtained. The effect from the parameters of space-time, the frequency of oscillator, the Cornell potential and the magnetic flux on the energy eigenvalues have been analyzed. We find a analogs effect for the bound states from the Aharonov-Bohm effect in our considered system.

\keywords{Generalized Duffin-Kemmer-Petiau oscillator; Som-Raychaudhuri space-time; Electromagnetic interactions; Aharonov-Bohm effects}
\end{abstract}

\ccode{PACS Nos.: 03.65.Pm, 11.27.+d.}

\section{Introduction}\label{sec:intro}
The relativistic DKP equation describe the fields and particles with spin-0 and spin-1, which has a similar algebraic structure with the Dirac equation \cite{duffin,non19,non22,castlb,hhdkp,non25}. Comparing the Klein-Gordon equations which is used to describe spin-0 particles and Proca equation which is used to describe spin-1 particles, the DKP equation is more universality. The DKP theory is extensively applied in various research direction, such as the research of quantum chromodynamics in large and short distances \cite{non45}, covariant Hamiltonian dynamics \cite{non50} and nuclear-hadron interactions \cite{REkoza}. In high-energy physics, the equation has been used to describe the interactions between the hadrons and nuclei \cite{Roger}.  Meanwhile, it has been extended to curved space-time \cite{non47,non48}.

The Dirac oscillator, Dirac equation with the linear harmonic potential, was firstly researched by Ito et al. \cite{Ito,Mosh}. Particulary, in their study they proposed that the momentum operator of free Dirac equation was substituted by $\vec{p} \rightarrow \vec{p}-i m \omega \beta \vec{r}$. As an exactly solvable model, Dirac oscillator have attracted much attention from theoretical physicists \cite{kk2,rll1,kk3}, especially the Dirac oscillator in non-commutative space \cite{vega}, high energy physics \cite{grin,muna} and non-relativistic supersymmetric quantum mechanics \cite{beck}. The DKP oscillator is similar with the Dirac oscillator \cite{boum49,boum346,cast75}, which is a sort of tensor coupling possessing the linear potential. Recently, theoretical physicists are increasingly interested in DKP oscillators, in particular in the background of cosmic strings and other topological defects \cite{Castro:2015zca,hh78293,Hosseinpour:2018udm,Hassanabadi:2018nvq,DeMontigny:2016tuu}. The conical space-time around the string result in a large amount of remarkable physical effects. For instance, due to the interaction between the atom and space-time curvature, the energy level of the atom will transform. So, we
must consider the space-time's topology to describe wholly the system's physics.

The G$\ddot{\mathrm{o}}$del-type space-times have cosmological solutions with rotating matter. Moreover, G$\ddot{\mathrm{o}}$del-type solutions demonstrate that closed time-like world-lines is permitted in general relativity. In Ref. \cite{metric}, the authors have introduced the G$\ddot{\mathrm{o}}$del-type metrics, which were used to gain the solutions of the all kinds of gravity and
super-gravity theories in disparate dimensions. In Ref. \cite{JDBa,RJG}, they have studied the G$\ddot{\mathrm{o}}$del-type's various aspects. From the G$\ddot{\mathrm{o}}$del-type space-times one can get the Som-Raychaudhuri space-time. Some theoretical physicists have studied numerous physical properties under the Som-Raychaudhuri space-time. For instance, spin-0 particles' relativistic quantum dynamics is affected by the this space-time\cite{wzepjp}, the linear confinement for the relativistic particle is studied in this space-time \cite{RLLconfine,2019yed}, the Dirac fermisions with torsion were studied under this space-time in Ref.  \cite{garcia} and the authors have studied the Dirac oscillator under the this space-time background \cite{monti}.

The generalization of the Klein-Gordon oscillator and Dirac oscillator has been studied\cite{2017hjk,202013520,dlfepjp2019,dlfahep,ahmed51}. Here, the ``generalization" is considered by the way that the momentum operator becomes $\vec{p} \rightarrow \vec{p}-i m \omega \beta \vec{r}$, where the radial coordinate $r$ is substituted by the potential function $f(r)$. On the other hand, in Refs. \cite{fa0,fa1} the authors have studied the Aharonov-Bohm effect on the spin-0 particle through investigating the generalized Klein-Gordon oscillator in curved space-times. In our work, we will introduce the generalized momentum operator to generalize DKP oscillator and analyze the Aharonov-Bohm effect for our considered problem under the Som-Raychaudhuri space-time.

The paper is planned as follows. In Sect. \ref{sec:two}, we will establish the generalized DKP oscillator under the Som-Raychaudhuri space-time. In Sect. \ref{sec:three}, we study the generalized DKP oscillator considered the Cornell potential under the Som-Raychaudhuri space-time and plot figures to analyze the effect from the parameter of space-time, the parameter of potential and the oscillator frequency on the energy eigenvalues. In Sect. \ref{sec:result}, the generalized DKP oscillator under the Som-Raychaudhuri space-time with the Aharonov-Bohm potential is solved. Meanwhile, we analyze the Aharonov-Bohm effect to the bound states. Eventually, the conclusions are given in Sect. \ref{sec:summary}.

\section{Generalized DKP oscillator under the Som-Raychaudhuri space-time}\label{sec:two}
The G$\ddot{\mathrm{o}}$del-type metric in cylindrical coordinates with torsion can be written as \cite{21447,jcar}
\begin{equation}\label{1}
d s^{2}=-\left(d t+\frac{\alpha \Omega \sinh ^{2} l r}{l^{2}} d \varphi\right)^{2}+d r^{2}+\alpha^{2} \frac{\sinh ^{2} 2 l r}{4 l^{2}} d \varphi^{2}+d z^{2},
\end{equation}
where $0 \leq r, 0 \leq \varphi \leq 2 \pi$, $-\infty<z<\infty$ and the parameter $\Omega > 0$ denotes the vorticity of the space-time. We use the natural units ($\hbar =1, c = 1$). The angular parameter is $\alpha=1-4m$, where $m$ is the linear mass density of the string and angular deficit has the ranges $0<\alpha<1$.

In our work, we focus on the Som-Raychaudhuri space-time, which is obtained when metric (\ref{1}) meeting the condition $l \rightarrow 0$. Hence, the line element can be read as \cite{monti}
\begin{equation}\label{linee}
d s^{2}=-\left(d t+\alpha \Omega r^{2} d \varphi\right)^{2}+d r^{2}+\alpha^{2} r^{2} d \varphi^{2}+d z^{2}.
\end{equation}

The researchers use usually similar manner that describe general-relativistic covariant Dirac equation to describe the dynamics of a DKP spinor under the curved space-time. The DKP equation under the curved space-time is given by \cite{lbca,hh541}
\begin{equation}\label{fdkp}
\left(i \beta^{\mu} \nabla_{\mu}-M\right) \Psi=0.
\end{equation}
The covariant derivative in Eq. (\ref{fdkp}) is given by
\begin{equation}
\nabla_{\mu}=\partial_{\mu}+\Gamma_{\mu}(x).
\end{equation}
Here, $\Gamma_{\mu}$ is the spinor connection which can be written as
\begin{equation}
\Gamma_{\mu}=\frac{1}{2} \omega_{\mu a b}\left[\beta^{a}, \beta^{b}\right],
\end{equation}
where beta-matrices meet the DKP algebra
\begin{equation}
\beta^{a} \beta^{c} \beta^{b}+\beta^{b} \beta^{c} \beta^{a}=\beta^{a} g^{c b}+\beta^{b} g^{c a},
\end{equation}
and the spin affine connection $\omega_{\mu a b}$ are given by
\begin{equation}\label{gmuij}
\omega_{\mu a b}=\eta_{a c} e_{v}^{c} e_{b}^{\sigma} \Gamma_{\sigma \mu}^{v}-\eta_{a c} e_{b}^{v} \partial_{\mu} e_{v}^{c},
\end{equation}
where $\eta_{a b}=\operatorname{diag}(-,+,+,+)$
is the Minkowski flat metric. Nevertheless,
$\beta^{\mu}$ in Eq. (\ref{fdkp}) is the DKP matrices in curved space-time, which are
obtained via
\begin{equation}
\beta^{\mu}(x)=e^{\mu}_{a}(x) \beta^{a}.
\end{equation}

According to the line element in Eq. (\ref{linee}), we can choose
following tetrads
\begin{equation}
e_{a}^{\mu}=\left(\begin{array}{cccc}1 & 0 & -\Omega r & 0 \\ 0 & 1 & 0 & 0 \\ 0 & 0 & \frac{1}{\alpha r} & 0 \\ 0 & 0 & 0 & 1\end{array}\right), \quad e^{a}_{\mu}=\left(\begin{array}{cccc}1 & 0 & \alpha \Omega r^{2} & 0 \\ 0 & 1 & 0 & 0 \\ 0 & 0 & \alpha r & 0 \\ 0 & 0 & 0 & 1\end{array}\right),
\end{equation}
which meet the orthonormality conditions
\begin{equation}\begin{array}{l}\displaystyle
e_{a}^{\mu}(x) e_{v}^{a}(x)=\delta_{v}^{\mu},\vspace{1.5ex} \\
\displaystyle e_{\mu}^{a}(x) e_{b}^{\mu}(x)=\delta_{b}^{a},
\end{array}\end{equation}
and the following relation
\begin{equation}
g_{\mu \nu}(x)=e_{\mu}^{a}(x) e_{v}^{b}(x) \eta_{a b}, \quad g^{\mu \nu}(x)=e_{a}^{\mu}(x) e_{b}^{v}(x) \eta^{a b}.
\end{equation}
The Christoffel symbols $\Gamma_{i j}^{\mu}$ in Eq. (\ref{gmuij}) are defined as
\begin{equation}
\Gamma_{i j}^{\mu}=\frac{1}{2} g^{\mu \nu}\left(g_{\nu i,j}+g_{\nu j,i}-g_{ij,\nu}\right).
\end{equation}
Therefore, $\beta^{\mu} \Gamma_{\mu}$ can be obtain
\begin{equation}
\beta^{\mu} \Gamma_{\mu}=\left(\begin{array}{ccccc}
0 & 0 & \displaystyle -\frac{1}{r} & 0 & 0 \\
0 & 0 & 0 & 0 & 0 \\
0 & 0 & 0 & 0 & 0 \\
0 & 0 & 0 & 0 & 0 \\
0 & 0 & 0 & 0 & 0
\end{array}\right).\end{equation}

The definition of the DKP oscillator is that
DKP equation has the Dirac oscillator interaction. Therefore, the DKP oscillator can be obtained from Eq. (\ref{fdkp}) by the nonminimal substitution
\begin{equation}i \nabla \rightarrow i \nabla-i M \omega \eta^{0} \mathbf{r},
\end{equation}
where $\omega$ represents the frequency of oscillator , $\eta^{0}=2\left(\beta^{0}\right)^{2}-1$ and $\mathbf{r} = (0, r, 0, 0)$. In our considered situation, the generalized oscillator is that the $r$ is replaced by $f(r)$. In other words, the vector $\mathbf{r}$ becomes $\mathbf{r} = (0, f(r), 0, 0)$.
Therefore, the
generalized DKP oscillator can be written as
\begin{equation}\label{generaldkp}
\left[i \beta^{\mu} \partial_{\mu}+i \beta^{\mu} \Gamma_{\mu}+i \beta^{r} M \omega \eta^{0} f(r)-M\right] \Psi(x)=0.
\end{equation}

Substituting the DKP spinor
\begin{equation}\label{ansatz}
\displaystyle\Psi(t, r, \varphi, z)=e^{-i E t+i m \varphi+i k z}\left(\begin{array}{c}\Phi_{1}(r) \\ \Phi_{2}(r) \\ \Phi_{3}(r) \\ \Phi_{4}(r) \\ \Phi_{5}(r)\end{array}\right)
\end{equation}
into Eq. (\ref{generaldkp}), we can arrive
the algebraic equations
\begin{equation}
\begin{aligned}
& -i r \alpha \frac{\mathrm{d}\Phi_{3}(r)}{\mathrm{d}r}+i \alpha\left(M \omega r f(r)-1\right) \Phi_{3}(r)+E r \alpha \Phi_{2}(r)
\vspace{1.5ex}\\&\quad+\left(m+E r^{2} \alpha \Omega\right) \Phi_{4}(r)+k r \alpha \Phi_{5}(r)-r \alpha M \Phi_{1}(r)=0, \vspace{1.5ex}\\ &E \Phi_{1}(r)-M \Phi_{2}(r)=0, \vspace{1.5ex}\\& i \frac{\mathrm{d}\Phi_{1}(r)}{\mathrm{d}r}+i M \omega f(r) \Phi_{1}(r)-M \Phi_{3}(r)=0, \vspace{1.5ex}\\& r^{2} \alpha E\Omega \Phi_{1}(r)+m\Phi_{1}(r)+r \alpha M \Phi_{4}(r)=0,\vspace{1.5ex} \\
&k \Phi_{1}(r)+M \Phi_{5}(r)=0.
\end{aligned}
\end{equation}
From above five equations, the equation of the Generalized DKP oscillator under the Som-Raychaudhuri space-time is given by
\begin{equation}\label{secondorder}
\begin{array}{l}\displaystyle\frac{\mathrm{d}^2}{\mathrm{d}r^2}\Phi_{1}(r)+\frac{1}{r} \frac{\mathrm{d}}{\mathrm{d}r}\Phi_{1}(r)-\left[M^{2}+k^{2}+M^2 \omega^2f^2(r) -\frac{M\omega f(r)}{r}\right. \vspace{1.5ex}\\\displaystyle \left.\quad+ E^{2}\left(r^{2} \Omega^{2}-1\right)-M \omega \frac{df(r)}{dr}+\frac{m^{2}}{r^{2} \alpha^{2}}+\frac{2 E m \Omega}{\alpha}\right] \Phi_{1}(r)=0.
\end{array}
\end{equation}
\section{The solution of the Generalized DKP oscillator}\label{sec:three}
In this section, we use the Nikiforov-Uvarov (NU) method \cite{NUmethod} to study the generalized DKP oscillator. The NU method has been used to investigate various physical background \cite{hh1,hh2,hh3,dsh1,dsh2,dsh3,dsh4, hh5,hh6,hh7,wz}. We make $f(r)$ become the Cornell potential, which include the linear term and Coulomb term. The linear term is a confining term, which can represents the non-perturbative effects of quantum chromodynamics. The second term is the non-confining term, which is caused through one-gluon swop among quark and antiquark. The Cornell potential plays an important role in particle physics. Let us set \cite{cp1,cp2,Chen:2020bqm,Ch1}
\begin{equation}f(r)=A r+\frac{B}{r},\end{equation}
where A and B denote the string tension and parameterizing of the Coulomb strength, respectively. By considered the Cornell potential into Eq. (\ref{secondorder}) we obtain
\begin{equation}
\begin{array}{l}\displaystyle\frac{\mathrm{d}^2\Phi(r)}{\mathrm{d}r^2}+\frac{1}{r}\frac{\mathrm{d}\Phi(r)}{\mathrm{d}r}-\left[\left(M^{2} \omega^{2}A^2+E^{2} \Omega^{2}\right) r^{2}+ \left(M^{2} \omega^{2}B^2+\frac{m^{2}}{\alpha^{2}}\right) \frac{1}{r^{2}}
\right.\vspace{1.5ex} \\\displaystyle \left.\quad-E^{2}+k^{2}+M^{2}-\frac{2 E m \Omega}{\alpha}-2 M \omega A+2ABM^2\omega^{2}\right] \Phi(r)=0.
\end{array}
\end{equation}
In order to obtain the solvable equation, we take the appropriate functional form
\begin{equation}
\displaystyle\Phi_{1}(r)=\frac{1}{\sqrt{r}} S(r).
\end{equation}
So we obtain
\begin{equation}\label{22}
\begin{array}{l}\displaystyle\frac{\mathrm{d}^2S(r)}{\mathrm{d}r^2}+\left[-\left(M^{2} \omega^{2}A^2+E^{2} \Omega^{2}\right) r^{2}+ \left(\frac{1}{4}-M^{2} \omega^{2}B^2-\frac{m^{2}}{\alpha^{2}}\right) \frac{1}{r^{2}}
\right.\vspace{1.5ex} \\\displaystyle \left.\quad+E^{2}-k^{2}-M^{2}-\frac{2 E m \Omega}{\alpha}+2 M \omega A-2ABM^2\omega^{2}\right] S(r)=0.
\end{array}
\end{equation}
When the new variables $\kappa = r^2$ is introduced, Eq. (\ref{22}) is transformed into
\begin{equation}\label{24}
\begin{array}{l}\displaystyle\frac{\mathrm{d}^2S(\kappa)}{\mathrm{d}\kappa^2}+\frac{1}{2\kappa}
\frac{\mathrm{d}S(\kappa)}{\mathrm{d}\kappa}+
\frac{1}{4\kappa^2}\left[-\left(M^{2} \omega^{2}A^2+E^{2} \Omega^{2}\right) \kappa^{2}+ \left(\frac{1}{4}-M^{2} \omega^{2}B^2-\frac{m^{2}}{\alpha^{2}}\right)
\right. \vspace{1.5ex} \\\displaystyle \left.\quad+\left(E^{2}-k^{2}-M^{2}-\frac{2 E m \Omega}{\alpha}+2 M \omega A-2ABM^2\omega^{2}\right)\kappa\right] S(\kappa)=0.
\end{array}
\end{equation}

We can see that the algebraic form of Eq. (\ref{24}) is similar to the  standard form of NU method.
Before giving the energy spectrum and wave function, we must calculate firstly the parameters of the NU method, which can be read as in our consider problem
\begin{equation}\begin{aligned}
\zeta_{1} &=\frac{1}{4}\left(M^{2} \omega^{2}A^2+E^{2} \Omega^{2}\right), \\
\zeta_{2} &=\frac{1}{4}\left(E^{2}-k^{2}-M^{2}-\frac{2 E m \Omega}{\alpha}+2 M \omega A-2ABM^2\omega^{2}\right), \\
\zeta_{3} &=\frac{1}{4}\left(\frac{m^{2}}{\alpha^{2}}-\frac{1}{4}+M^{2} \omega^{2}B^2\right), \\
c_{1} &=\frac{1}{2}, \quad c_{2} =c_{3}=0, \quad c_{4} =\frac{1}{2},\quad c_{5}=0, \\
c_{6} &=\frac{1}{4}\left(M^{2} \omega^{2}A^2+E^{2} \Omega^{2}\right), \\
c_{7} &=-\frac{1}{4}\left(E^{2}-k^{2}-M^{2}-\frac{2 E m \Omega}{\alpha}+2 M \omega A-2ABM^2\omega^{2}\right),\\
c_{8} &=\frac{1}{4}\left(\frac{m^{2}}{\alpha^{2}}-\frac{1}{4}+M^{2} \omega^{2}B^2\right)+\frac{1}{16}, \\
c_{9} &=\frac{1}{4}\left(M^{2} \omega^{2}A^2+E^{2} \Omega^{2}\right), \\
c_{10} &=1+2\sqrt{\frac{1}{16}+\frac{1}{4}\left(\frac{m^{2}}{\alpha^{2}}-\frac{1}{4}+M^{2} \omega^{2}B^2\right)},\\
c_{11} &=2\sqrt{\frac{1}{4}\left(M^{2} \omega^{2}A^2+E^{2} \Omega^{2}\right)}, \\
c_{12} &=\frac{1}{4}+\sqrt{\frac{1}{16}+\frac{1}{4}\left(\frac{m^{2}}{\alpha^{2}}-\frac{1}{4}+M^{2} \omega^{2}B^2\right)}, \\
c_{13} &=-\sqrt{\frac{1}{4}\left(M^{2} \omega^{2}A^2+E^{2} \Omega^{2}\right)}.
\end{aligned}\end{equation}
According to the NU method, through the above parameters we can get the eigenvalues
\begin{equation}\label{ES}
\begin{aligned}
&2(2n+1)\sqrt{M^{2} \omega^{2}A^2+E^{2} \Omega^{2}}-\left(E^{2}-k^{2}-M^{2}-\frac{2 E m \Omega}{\alpha}+2 M \omega A-2ABM^2\omega^{2}\right)\\
&+2\sqrt{\left(M^{2} \omega^{2}A^2+E^{2} \Omega^{2}\right)\left(M^{2} \omega^{2}B^2+\frac{m^2}{\alpha^2}\right)}=0,
\end{aligned}
\end{equation}
and the eigenfunction of generalized DKP oscillator under the Som-Raychaudhuri space-time
\begin{equation}
\begin{aligned}
\Phi_1(r)= r^{\Theta}\text{exp}\left(-\frac{1}{2}\sqrt{M^{2} \omega^{2}A^2+E^{2} \Omega^{2}}r^2\right)L^{\Theta}_n\left(\sqrt{M^{2} \omega^{2}A^2+E^{2} \Omega^{2}}r^2 \right),
\end{aligned}
\end{equation}
where $\displaystyle\Theta=\sqrt{M^{2} \omega^{2}B^2+\frac{m^2}{\alpha^2}}$, and $L$ is generalized Laguerre polynomial.

Meanwhile, the charge density is given by
\begin{equation}
\begin{aligned}
J^{t} \propto \bar{\Psi} \beta^{t} \Psi=\Psi^{\dagger} \eta^{0} \beta^{t} \Psi=-2 \frac{E\alpha\left(r^{2} \Omega^{2}-1\right)+m \Omega}{M \alpha}\left|\Phi_{1}(r)\right|^{2}.
\end{aligned}
\end{equation}

\begin{figure}[b!]
\vspace{-1.0cm}
\begin{center}
\includegraphics[width=4.5in,height=3.2in]{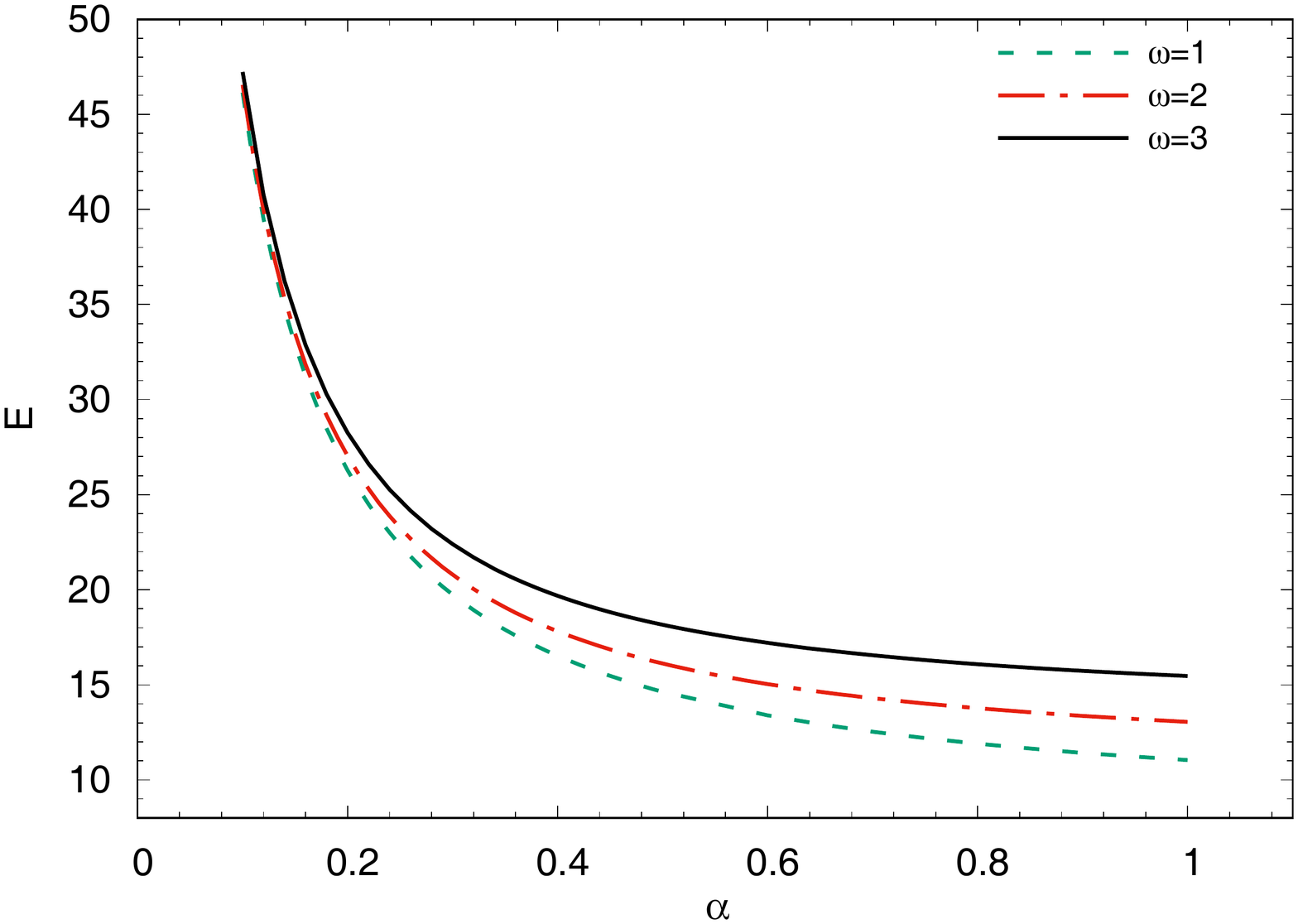}
\end{center}
\setlength{\abovecaptionskip}{-1.0cm}
\setlength{\belowcaptionskip}{0.5cm}
\caption{Energy eigenfunctions from Eq. (\ref{ES}) as a function of $\alpha$ according to three different values of $\omega$ for $n=m=k=M=\Omega=A=B=1$.}
\label{Ea}
\vspace{0.1cm}
\end{figure}
\begin{figure}[t!]
\vspace{-1.3cm}
\begin{center}
\includegraphics[width=4.6in,height=3.2in]{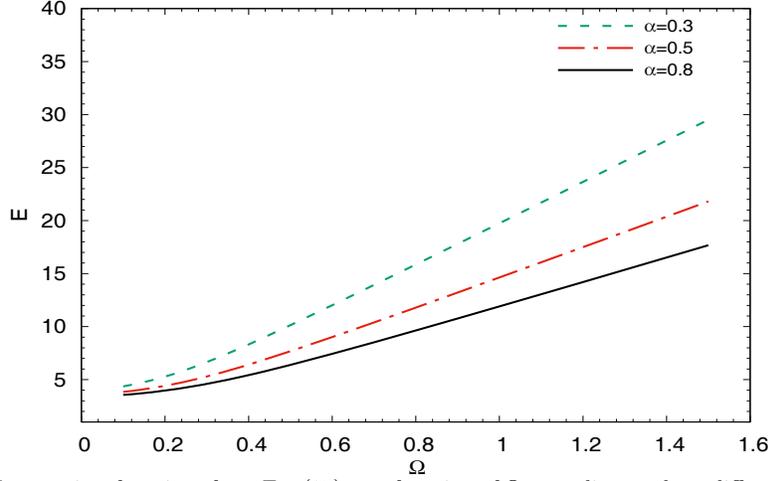}
\end{center}
\setlength{\abovecaptionskip}{-1.2cm}
\setlength{\belowcaptionskip}{0.5cm}
\caption{Energy eigenfunctions from Eq. (\ref{ES}) as a function of $\Omega$ according to three different values of $\alpha$ for $n=m=k=M=\omega=A=B=1$.}
\label{EO}
\vspace{0.1cm}
\end{figure}
\begin{figure}[t!]
\vspace{-1.0cm}
\begin{center}
\includegraphics[width=4.6in,height=3.2in]{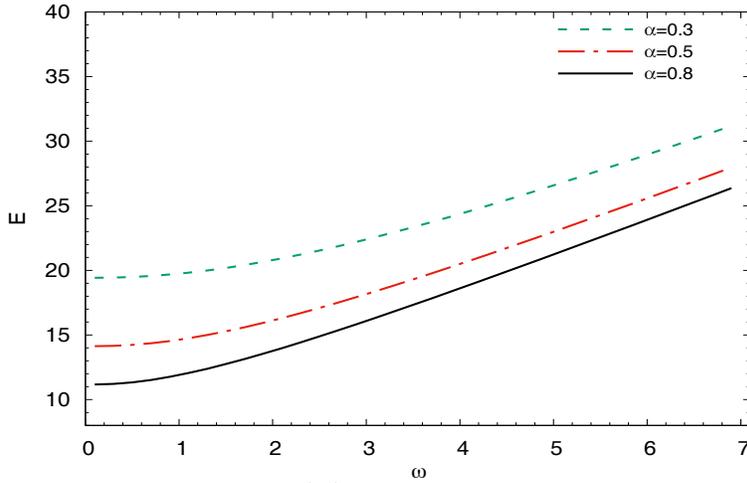}
\end{center}
\setlength{\abovecaptionskip}{-1.2cm}
\setlength{\belowcaptionskip}{0.5cm}
\caption{Energy eigenfunctions from Eq. (\ref{ES}) as a function of $\omega$ according to three different values of $\omega$ for $n=m=k=M=\Omega=A=B=1$.}
\label{Ew}
\vspace{0.2cm}
\end{figure}
\begin{figure}[t!]
\vspace{-1.0cm}
\begin{center}
\includegraphics[width=4.6in,height=3.2in]{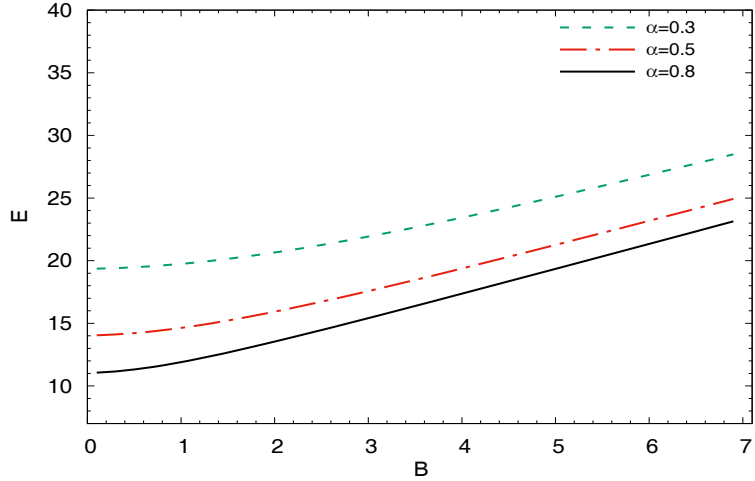}
\end{center}
\setlength{\abovecaptionskip}{-1.0cm}
\setlength{\belowcaptionskip}{0.5cm}
\caption{Energy eigenfunctions from Eq. (\ref{ES}) as a function of $\omega$ according to three different values of $\omega$ for $n=m=k=M=\omega=\Omega=A=1$.}
\label{EB}
\vspace{0.2cm}
\end{figure}

Now, we analyze the affect of all parameters on the generalized DKP oscillator with the Cornell potential. We only consider the positive particles corresponding to the positive energy eigenvalues. In Fig. \ref{Ea}, the energy spectrum distribution of the generalized DKP oscillator with the Cornell potential as a function of $\alpha$ is given, when $n=m=k=M=\Omega=A=B=1$. From Fig. \ref{Ea}, we can obtain the information that the energy spectrum of the spin-0 particles sharply decrease with $\alpha$ in small $\alpha$ region, while the trend is weaker in large $\alpha$ region. In Fig. \ref{EO}, the energy spectrum distribution of the generalized DKP oscillator with the Cornell potential as a function of $\Omega$ is given, when $n=m=k=M=\omega=A=B=1$. From Fig. \ref{EO}, one can see that the energy spectrum of the spin-0 particles increase with vorticity parameter $\Omega$. In Fig. \ref{Ew}, the energy spectrum distribution of the generalised DKP oscillator with the Cornell potential as a function of $\omega$ is given, when $n=m=k=M=\Omega=A=B=1$. From Fig. \ref{Ew}, one can see that the energy spectrum of the spin-0 bosons increase with $\omega$. In Fig. \ref{EB}, the energy spectrum distribution of the generalized DKP oscillator with the Cornell potential as a function of $B$ is given, when $n=m=k=M=\omega=\Omega=A=1$. From Fig. \ref{EB}, one can see that the eigenvalues of the spin-0 particles increase slowly with Cornell potential parameters $B$.

\section{Generalized DKP oscillator with the
electromagnetic potential} \label{sec:result}
We study the generalized DKP oscillator with the electromagnetic potential in this section. As a quantum mechanical phenomenon, Aharonov-Bohm effect is that charged particle will be effected by the electromagnetic potential although limited to the areas where don't exist magnet field. The Aharonov-Bohm potential has been widely discussed in curved space-time \cite{fa2,fa3,fa4}. In zero magnetic field region, the charged particle move along path $P$. This particle obtain a phase shift
\begin{equation}\varphi=e \int_{P} \vec{A} \cdot d \vec{x}.
\end{equation}
Therefore, when the particles have the same starting point and ending point but the paths are different, a phase difference will produce
\begin{equation}\triangle \varphi=e \Phi_{B}.
\end{equation}
In our work, we consider the electromagnetic interactions \cite{ele} into the Generalized DKP oscillator equation by using the minimal substitution, $p_{\mu} \rightarrow p_{\mu}-e A_{\mu}$, which is  \cite{abp}
\begin{equation}\partial_{\mu} \rightarrow \partial_{\mu}-i e A_{\mu}.
\end{equation}
In this work, we choose Aharonov-Bohm potential, which is given by
\begin{equation}e A_{\phi}=\Phi \quad, \quad \Phi=\frac{\Phi_{B}}{(2 \pi / e)},
\end{equation}
where $\Phi_{B}$ represents magnetic quantum flux.

Thereby, based on the Eq. (\ref{generaldkp}) the generalized DKP oscillator considered the electromagnetic interactions can be written as
\begin{equation}\label{magneticDKP}
\left[i \beta^{t} \partial_{t}+i \beta^{r}\left(\partial_{r}+M \omega \eta^{0}f(r)\right)+i \beta^{\phi}\left(\partial_{\phi}-i e A_{\phi}\right)+i \beta^{z}\partial_{z}+i \beta^{\mu} \Gamma_{\mu}-M\right] \Psi=0.
\end{equation}

Substituting the Eq. (\ref{ansatz}) into the Eq. (\ref{magneticDKP}), we get
\begin{equation}\label{33}
\begin{aligned}
& -i \alpha r \frac{\mathrm{d}\Phi_{3}(r)}{\mathrm{d}r}+i \alpha\left(M \omega r f(r)-1\right) \Phi_{3}(r)+E r \alpha \Phi_{2}(r)
\vspace{1.5ex}\\&\quad+\left(m+E r^{2} \alpha \Omega\right) \Phi_{4}(r)+k r \alpha \Phi_{5}(r)-r \alpha M \Phi_{1}(r)-\frac{e\Phi_{B}}{2\pi}\Phi_{4}(r)=0, \vspace{1.9ex}\\
&E \Phi_{1}(r)-M \Phi_{2}(r)=0, \vspace{1.5ex}\\&
i\frac{\mathrm{d}\Phi_{1}(r)}{\mathrm{d}r}+i M \omega f(r) \Phi_{1}(r)-M \Phi_{3}(r)=0, \vspace{1.5ex}\\
&r^{2} \alpha E\Omega \Phi_{1}(r)+m\Phi_{1}(r)+r \alpha M \Phi_{4}(r)-\frac{e\Phi_{B}}{2\pi}\Phi_{1}(r)=0\vspace{1.5ex} \\
&k \Phi_{1}(r)+M \Phi_{5}(r)=0.
\end{aligned}
\end{equation}
From above equations, we obtain
\begin{equation}\label{decouple}
\begin{aligned}
&\Phi_{2}(r)=\frac{E \Phi_{1}(r)}{M}, \vspace{1.5ex}\\
&\Phi_{3}(r)=\frac{i\displaystyle\frac{\mathrm{d}\Phi_{1}(r)}{\mathrm{d}r}+i M \omega f(r) \Phi_{1}(r)}{M}, \vspace{1.5ex}\\
&\Phi_{4}(r)=\frac{-r^{2} \alpha E\Omega \Phi_{1}(r)-m\Phi_{1}(r)+\displaystyle\frac{e\Phi_{B}}{2\pi}\Phi_{1}(r)}{r \alpha M }, \\
&\Phi_{5}(r)=\frac{-k \Phi_{1}(r)}{M}.
\end{aligned}
\end{equation}
By substituting Eq. (\ref{decouple}) into the first equation of Eq. (\ref{33}), the radial equation considered the electromagnetic interactions is obtained
\begin{equation}\label{35}
\begin{aligned}
& \frac{\mathrm{d}^2}{\mathrm{d}r^2}\Phi_{1}(r)+\frac{1}{r} \frac{\mathrm{d}}{\mathrm{d}r}\Phi_{1}(r)-\left[M^{2}+k^{2}+M^2 \omega^2f^2(r) -\frac{M\omega f(r)}{r}+ E^{2}+E^{2}r^{2} \Omega^{2}\right. \\
&\left. -M \omega \frac{df(r)}{dr} +\frac{m^{2}}{r^{2} \alpha^{2}}+\frac{2 E m \Omega}{\alpha}-\frac{me\Phi_B}{\pi\alpha^2r^2}+
\frac{e^2\Phi_B^2}{4\pi^2\alpha^2r^2}-\frac{\Omega Ee\Phi_B}{\pi\alpha}\right] \Phi_{1}(r)=0.
\end{aligned}
\end{equation}
By substituting the Cornell potential into Eq. (\ref{35}) we obtain
\begin{equation}\label{36}
\begin{array}{l}\displaystyle\frac{\mathrm{d}^2\Phi(r)}{\mathrm{d}r^2}+\frac{1}{r}\frac{\mathrm{d}\Phi(r)}{\mathrm{d}r}-\left[\left(M^{2} \omega^{2}A^2+E^{2} \Omega^{2}\right) r^{2}+ \left(M^{2} \omega^{2}B^2+\frac{m^{2}}{\alpha^{2}}-\frac{e\Phi_B m}{\pi\alpha^{2}}+\frac{e^2 \Phi^{2}_B}{4\pi^2\alpha^{2}} \right)\frac{1}{r^{2}}
\right.\vspace{1.5ex} \\\displaystyle \left.\quad-E^{2}+k^{2}+M^{2}-\frac{2 E m \Omega}{\alpha}-2 M \omega A+2ABM^2\omega^{2}-\frac{\Omega E e\Phi_B}{\pi\alpha}\right] \Phi(r)=0.
\end{array}
\end{equation}
Let us define $Q(r )$ as
\begin{equation}
\displaystyle\Phi_{1}(r)=\frac{1}{\sqrt{r}} Q(r).
\end{equation}
So Eq. (\ref{36}) becomes
\begin{equation}\label{38}
\begin{array}{l}\displaystyle\frac{\mathrm{d}^2Q(r)}{\mathrm{d}r^2}+\left[-\left(M^{2} \omega^{2}A^2+E^{2} \Omega^{2}\right) r^{2}+ \left(\frac{1}{4}-M^{2} \omega^{2}B^2-\frac{m^{2}}{\alpha^{2}}+\frac{e\Phi_B m}{\pi\alpha^{2}}-\frac{e^2 \Phi^{2}_B}{4\pi^2\alpha^{2}}\right) \frac{1}{r^{2}}
\right.\vspace{1.5ex} \\\displaystyle \left.\quad+E^{2}-k^{2}-M^{2}-\frac{2 E m \Omega}{\alpha}+2 M \omega A-2ABM^2\omega^{2}+\frac{\Omega E e\Phi_B}{\pi\alpha}\right] Q(r)=0.
\end{array}
\end{equation}
When we consider the new variables $r^2 =\tau$, Eq. (\ref{38}) takes the form
\begin{equation}
\begin{array}{l}
\displaystyle\frac{\mathrm{d}^2Q(\tau)}{\mathrm{d}\tau^2}
+\frac{1}{2\tau}
\frac{\mathrm{d}Q(\tau)}{\mathrm{d}\tau}+
\frac{1}{4\tau^2}\left[-\left(M^{2} \omega^{2}A^2+E^{2} \Omega^{2}\right) \tau^{2}+ \frac{1}{4}-M^{2} \omega^{2}B^2-\frac{m^{2}}{\alpha^{2}}+\frac{e\Phi_B m}{\pi\alpha^{2}}
\right. \vspace{1.5ex} \\
\displaystyle \left.-\frac{e^2 \Phi^{2}_B}{4\pi^2\alpha^{2}}+\left(E^{2}-k^{2}-M^{2}-\frac{2 E m \Omega}{\alpha}+2 M \omega A-2ABM^2\omega^{2}+\frac{\Omega E e\Phi_B}{\pi\alpha}\right)\tau\right] Q(\tau)=0.
\end{array}
\end{equation}
In order to simplify above equation, the following variables are defined
\begin{equation}\label{40}
\begin{aligned}
\Lambda_1 &=-\left(M^{2} \omega^{2}A^2+E^{2} \Omega^{2}\right), \\
\Lambda_2 &=\frac{1}{4}-M^{2} \omega^{2}B^2-\frac{m^{2}}{\alpha^{2}}+\frac{e\Phi_B m}{\pi\alpha^{2}}-\frac{e^2 \Phi^{2}_B}{4\pi^2\alpha^{2}}, \\
\Lambda_3 &=E^{2}-k^{2}-M^{2}-\frac{2 E m \Omega}{\alpha}+2 M \omega A-2ABM^2\omega^{2}+\frac{\Omega Ee\Phi_B}{\pi\alpha}. \end{aligned}
\end{equation}
Therefore, we can obtain
\begin{equation}\label{41}
\displaystyle\frac{\mathrm{d}^2Q(\tau)}{\mathrm{d}\tau^2}
+\frac{1}{2\tau}
\frac{\mathrm{d}Q(\tau)}{\mathrm{d}\tau}+
\frac{1}{4\tau^2}\left(\Lambda_1 \tau^{2}+\Lambda_2+\Lambda_3\tau\right) Q(\tau)=0.
\end{equation}
According to the NU method's provisions, the eigenvalues considered the electromagnetic interactions becomes
\begin{equation}\label{42}
\begin{aligned}
2(2n+1)\sqrt{-\frac{\Lambda_1}{4}}-\frac{\Lambda_3}{4}+
2\sqrt{-\frac{\Lambda_1}{4}\left(\frac{1}{16}-\frac{\Lambda_2}{4} \right)}=0,
\end{aligned}
\end{equation}
Substituting the Eq. (\ref{40}) into (\ref{42}), we can arrive
\begin{equation}\label{43}
\begin{aligned}
&2(2n+1)\sqrt{M^{2} \omega^{2}A^2+E^{2} \Omega^{2}}-\left(E^{2}-k^{2}-M^{2}-\frac{2 E m \Omega}{\alpha}+2 M \omega A-2ABM^2\omega^{2}+\frac{\Omega Ee\Phi_B}{\pi\alpha}\right)\\
&+2\sqrt{\left(M^{2} \omega^{2}A^2+E^{2} \Omega^{2}\right)\displaystyle\left(M^{2} \omega^{2}B^2+\frac{m^2}{\alpha^2}-\frac{e\Phi_B m}{\pi\alpha^{2}}+\frac{e^2 \Phi^{2}_B}{4\pi^2\alpha^{2}}\right)}=0.
\end{aligned}
\end{equation}
Meanwhile, the eigenfunction becomes
\begin{equation}
\begin{aligned}
\Phi(r)=r^{\mathbb{C}}\text{exp}\left(-\frac{1}{2}\sqrt{M^{2} \omega^{2}A^2+E^{2} \Omega^{2}}r^2\right) L^{\mathbb{C}}_n\left(\sqrt{M^{2} \omega^{2}A^2+E^{2} \Omega^{2}}r^2 \right),
\end{aligned}
\end{equation}
where the parameter $\mathbb{C}=\displaystyle\sqrt{M^{2} \omega^{2}B^2+\frac{m^2}{\alpha^2}-\frac{e\Phi_B m}{\pi\alpha^{2}}+\frac{e^2 \Phi^{2}_B}{4\pi^2\alpha^{2}}}$.

From Eq. (\ref{43}), one can observe the relativistic energy eigenvalues of the generalized DKP oscillator have a dependence on the Aharonov-Bohm quantum phase. Therefore, we can get the relation $E_{n, l}\left(\Phi_{B}+\Phi_{0}\right)=E_{n, l \mp \epsilon}\left(\Phi_{B}\right)$, where $\Phi_{0}=\pm \frac{2 \pi}{e} \epsilon$ ($\epsilon=1,2,3, \cdots$). The energy spectrum of the generalized DKP oscillator rely on the geometric quantum phase, which bring about similar impact from the Aharonov-Bohm effect to the bound states \cite{kk1,kkab}.
\section{Conclusions} \label{sec:summary}
In this work, we have investigated generalized DKP oscillator for spin-0 bosons under the Som-Raychaudhuri space-time with a topological defect, including the cases of without and with the electromagnetic interactions. The equation of the generalized DKP oscillator under the Som-Raychaudhuri space-time background are derived. By using the NU method, we have calculated the eigenfunction and energy eigenvalues, and plotted the figures of energy eigenvalues as the function of space-time parameters, the frequency of oscillator and potential parameters. The effect of angular deficit $\alpha$, the vorticity parameter $\Omega$, the oscillator frequency $\omega$ and potential parameters on the energy spectrum are analyzed . On the other hand, by means of the minimal
substitution into the generalized DKP oscillator, the generalized DKP oscillator with electromagnetic interaction is introduced. We obtained the wave equation having the Aharonov-Bohm potential under the Som-Raychaudhuri space-time background. Meanwhile, the eigenfunction and energy eigenvalues are also obtained in this case. Our result shows a dependence the eigenfunction and energy spectrum about the Aharonov-Bohm magnetic flux. Especially,
the relativistic energy eigenvalues rely on the geometric quantum phase, which bring about similar impact for the bound states from the Aharonov-Bohm effect.

\section*{Acknowledgments}
This work was supported by the National Natural Science Foundation of China (Grant No. 11465006 and 11565009); and the Major Research Project of innovative Group of Guizhou province (2018-013).

\end{document}